\documentclass[prb,twocolumn]{revtex4-1}
\usepackage{amsmath}
\usepackage{amsfonts}
\usepackage{stmaryrd}
\usepackage{amssymb}
\usepackage{graphics,epsfig}
\usepackage{subfigure}
\usepackage{bm}
\usepackage{mathpple}
\input epsf.sty

\begin{document}
\title{Strain induced superconducting pair-density-wave states in graphene}
\author{Feng Xu$^{1}$, Po-Hao Chou$^{2}$, Chung-Hou Chung$^{3}$, Ting-Kuo Lee$^{4}$,  and Chung-Yu Mou $^{2,4,5}$}
\affiliation{$^{1}$School of Physics and Telecommunication Engineering,
Shaanxi University of technology, Hanzhong 723001, China}
\affiliation{$^{2}$Center for Quantum Technology and Department of Physics, National Tsing Hua University, Hsinchu 30043,
Taiwan, 300, R.O.C.}

\affiliation{$^{3}$Electrophysics Department, National Chiao-Tung University, HsinChu, Taiwan, R.O.C.}
\affiliation{$^{4}$Institute of Physics, Academia Sinica, Nankang 115, Taiwan, Republic of China}
\affiliation{$^{5}$Physics Division, National Center for Theoretical Sciences, Hsinchu 30013, Taiwan, R.O.C.}

\begin{abstract}
Graphene is known to be non-superconducting. However, surprising superconductivity is recently discovered in a flat-band in a twisted bi-layer graphene. Here we show that superconductivity can be more easily realized in topological flat-bands induced by strain in graphene through periodic ripples. Specifically, it is shown that by including correlation effects, the chiral d-wave superconductivity can be stabilized under strain even for slightly doped graphene. The chiral d-wave superconductivity generally coexists with charge density waves  (CDW) and pair density waves (PDW) of the same period. Remarkably, a pure PDW state with doubled period that coexists with the CDW state is found to 
emerge at  a finite temperature region under reasonable strain strength. The emergent PDW state is shown to be superconducting 
with non-vanishing superfluid density, and it realizes the long searched superconducting states with non-vanishing center of mass 
momentum for Cooper pairs.
\end{abstract}
\maketitle
\section{Introduction}
The issue of what alternative forms of superconducting states other than the BCS superconducting states can be realized
has been one of the main drives for searching unconventional superconductivity in condensed matter. 
In the high temperature superconductivity discovered in cuprates, it is now widely accepted that both the mechanism and the pairing symmetry are different from those in the conventional superconductivity\cite{highTc}. Furthermore, while the Cooper pairs have zero center-of-mass momentum in the conventional superconductivity and the charge density waves (CDW) are usually considered as being incompatible with this property\cite{BCS}, it is also realized that both CDW and pair density waves (PDW) that break translational symmetry are  intertwined and can even coexist with the superconducting order\cite{highTc,intertwined, PALee}.
More recently, it is put forth that while in conventional superconductors, the critical temperature is limited by
the Debye frequency $\omega_D$  through the relation for critical temperature $k_BT_c = \hbar \omega_D e^{-1/Ng}$, 
in an extreme limit when the electronic band is dispersion-less and becomes a flat band, the divergence of the density of states $N$ 
near the Fermi energy leads to enhanced critical temperature that is in proportional to the electron-phonon coupling constant $g$, i.e., 
$k_BT_c=g/2$\cite{flatbandSC}. The flat-band superconductivity is based on naive extrapolation of the BCS theory. In real materials, 
however, decreasing electronic bandwidth enhances on-site Coulomb interaction and may induce other instabilities such as 
CDW, antiferromagnetic order, ferromagnetism\cite{ferro} and etc. 
Indeed, as-grown graphene is known to be non-superconducting. However, in a recent experiment, superconductivity with strong correlation effects is discovered in a flat band arising in a slightly-twisted bilayer graphene\cite{grapheneSC,correlation_graphene}.  The discovered flat-band superconductivity indicates that graphene may host unconventional superconductivity under appropriate conditions.

In this paper, we explore superconducting phases in flat bands formed by an alternative way in graphene.
Unlike the flat-band in twisted bilayer graphene that requires fine tuning of the twisted angle, 
here flat bands are formed topologically by strain and can be robustly induced as Landau levels due to the
corresponding pseudo-magnetic field generated by the strain\cite{strain}. 
Experimentally, flat-bands in strained graphene have been observed with the strain being imposed or engineered 
by external stretching or periodic ripples\cite{strain, strain2}.  
Here by including correlation effects in graphene under periodic strain,  
it is shown that unconventional superconducting states with chiral d-wave symmetry can be stabilized 
even in slightly doped graphene. Furthermore, due to the periodicity introduced by strain, we find that the 
chiral d-wave superconductivity generally coexists with CDW and PDW of the same period. 
Remarkably, a pure PDW state with doubled period that coexists with CDW is found to 
emerge at  a finite temperature region under reasonable strain strength. The emergent PDW state is shown to be superconducting 
with non-vanishing superfluid density and realizes the long searched superconducting states with non-vanishing center of mass 
momentum for Cooper pairs\cite{FFLO}.

\section{Theoretical Model and Results}
\begin{figure}[t]
\includegraphics[height=1.3in,width=1.5in] {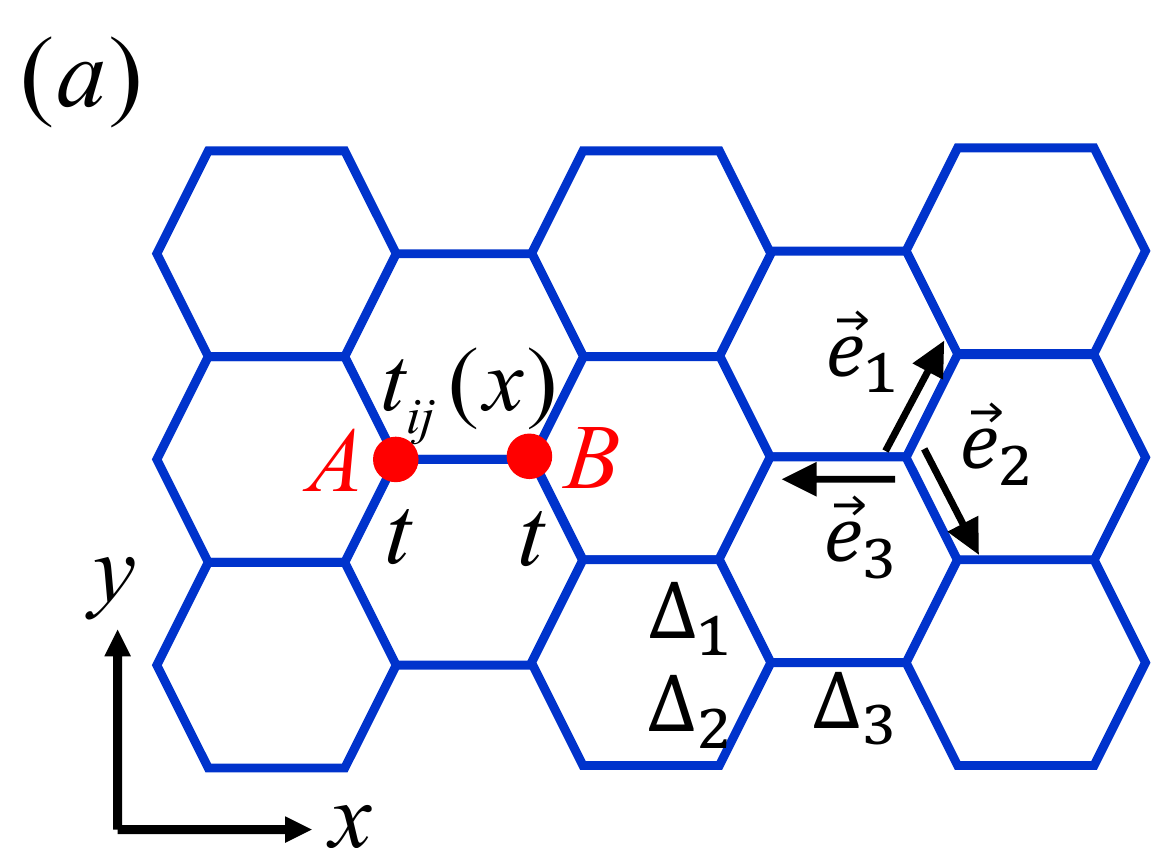}
\includegraphics[height=1.3in,width=1.3in] {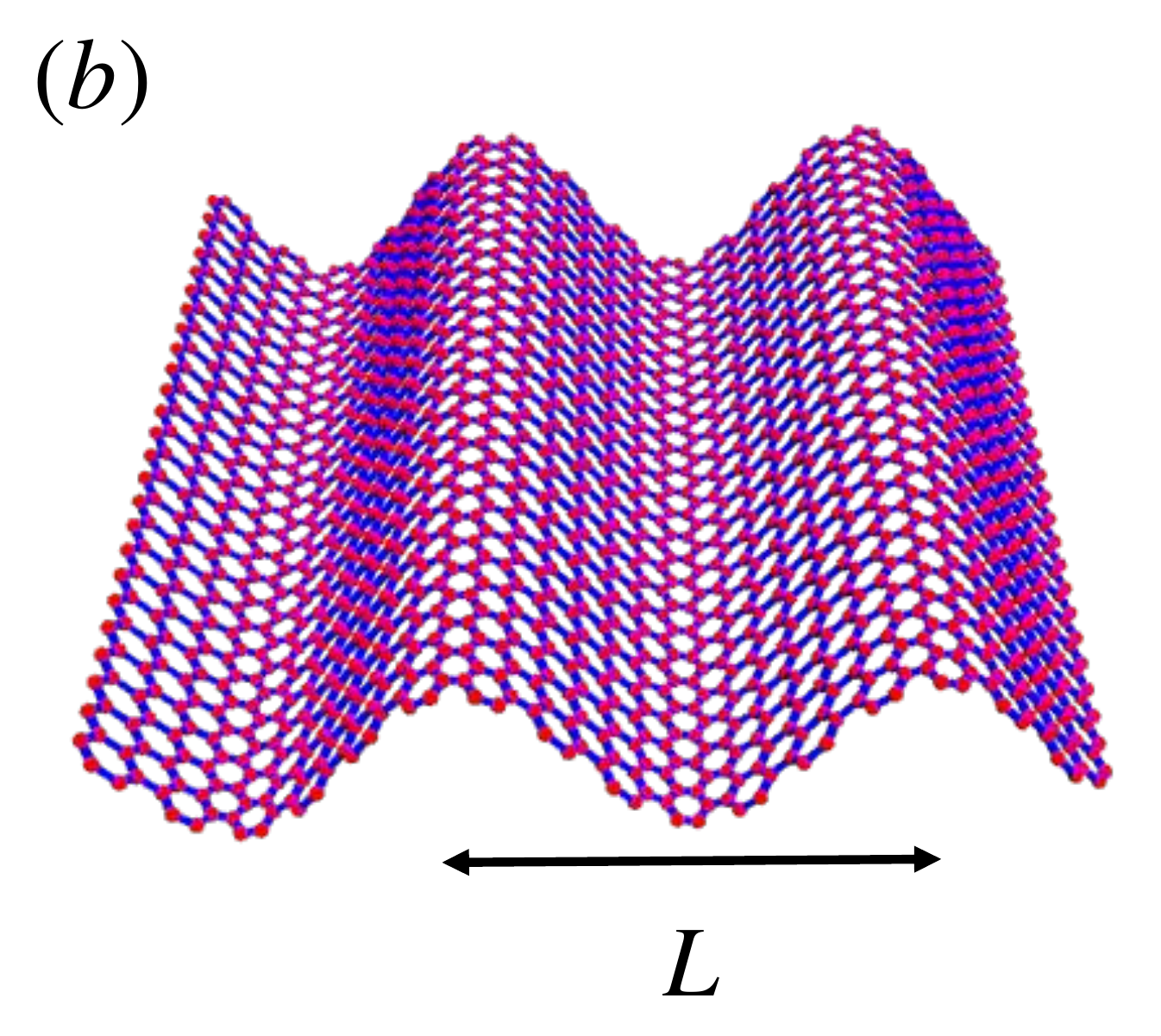}
\includegraphics[height=1.6in,width=2.8in] {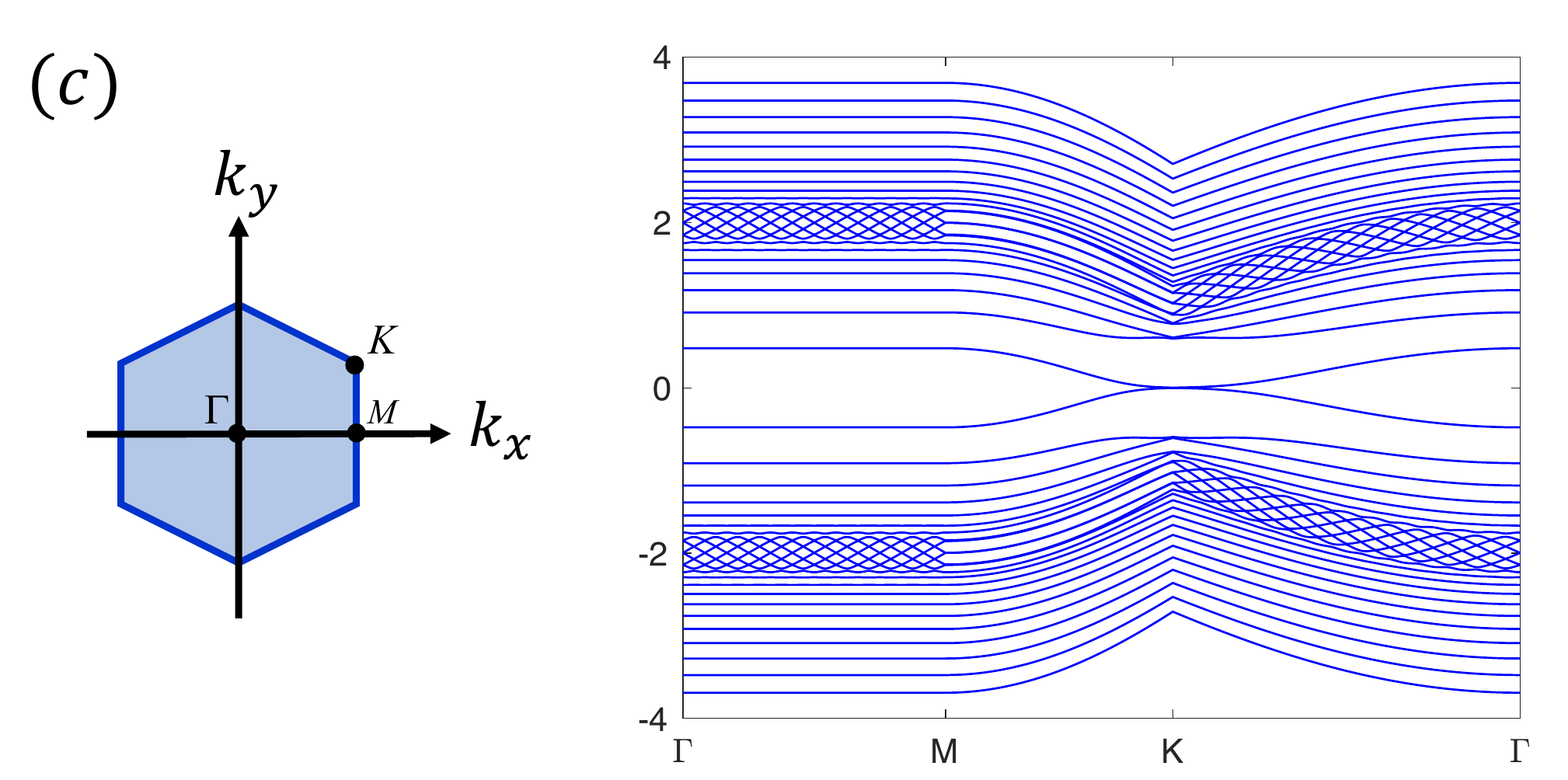}
\includegraphics[height=2.4in,width=3.0in] {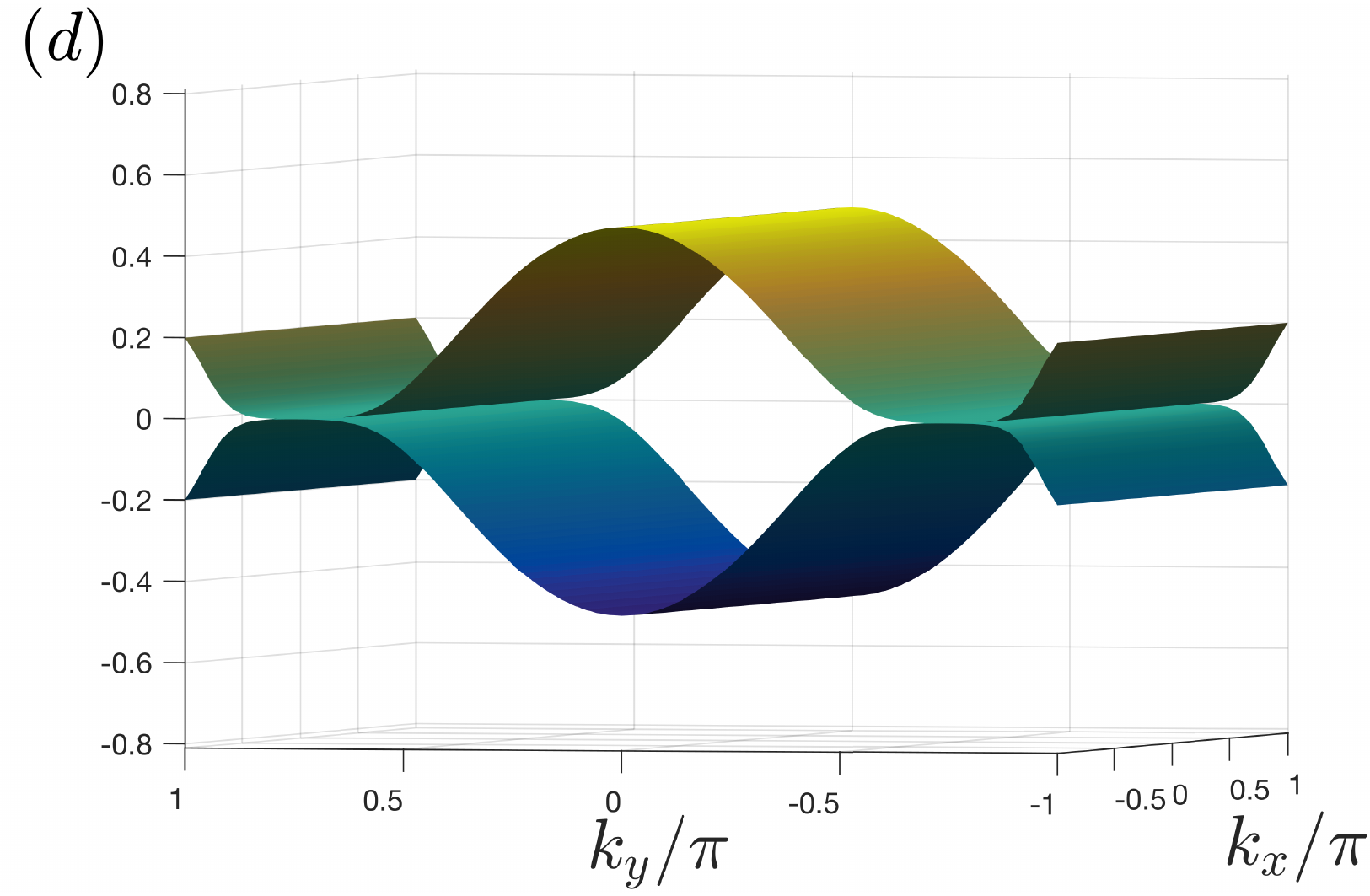}
\caption{Topological flat bands in strained graphene (a) Strain induced change of hopping amplitude. Here $t_{ij}(x) = t[1+\alpha \cos (Q x_i)]$ with $i$ denoting site $A$ and $j$ denoting site $B$. $\Delta_i$ represent three superconducting pairing amplitudes. (b) Schematic plot of the ripple with period $L$ (c) Plot of energy bands for $\alpha = 0.8$ and $L=24$ in the Brillouin zone of unstrained graphene. (d) Extension of zero energy flat -band (along $M$-$K$ direction) in strained graphene with $\alpha = 0.8$ and $L=24$}
\label{fig1}
\end{figure}

We start by considering the graphene under periodic strain.  As shown in Fig.~\ref{fig1}(b), the strain can be induced by ripple with fixed period $L$ or by external stretching. The strain generally induces changes of hopping amplitudes $t$  through the change of bond lengths $\delta_1$, $\delta_2$ and $\delta_3$ as $t_i = t \exp [-3.37 (|\vec{e}_i|/a-1) ]$\cite{hopping}(See Fig.~\ref{fig1}(a)). Here $t \approx 2.8$ eV is the equilibrium hopping amplitude, $a = 1.42 \AA$ is the equilibrium bond length, and $\vec{e}_i$ are three deformed nearest-neighbor
vectors whose corresponding undeformed vectors are $\vec{e}^0_1=\frac{a}{2}(1,\sqrt{3})$,~$\vec{e}^0_2=\frac{a}{2}(1,-\sqrt{3})$, and $\vec{e}^0_{3}=-a(1,0)$. In the simplest realization, we shall keep $e_1$ and $e_2$ fixed and deform $e_3$ with the period $L$\cite{simple}. The corresponding change in the hopping amplitude along the horizontal bond is given by
\begin{equation}
t_{ij} = t[1+\alpha \cos (Qx_i)], \label{tx}
\end{equation}
where $x$ labels the position of the left-hand site ($A$ in Fig.~\ref{fig1}(a)) of the bond $AB$ and the $Q=2\pi/L$ is the wavevector associated with the strain.  For ripples with the wavelength being in the nano-meter regime, $0<\alpha \le 0.5$ and $L=0.1-10$nm\cite{strain}.

The tight-binging Hamiltonian under strain is given by
\begin{equation}
H_0=-\sum_{i,j=1,2,\sigma} t c^{\dagger}_{i,\sigma}c_{i+\vec{e}_j,\sigma}-\sum_{i,\sigma} t(x_i)  c^{\dagger}_{i,\sigma}c_{i+\vec{e}_3,\sigma}+h.c.,
\end{equation}
where $i$ labels sub-lattice A, $t(x_i)=t_{i,i +\vec{e}_3}$, and $c_{i,\sigma}$ annihilates an electron with spin $\sigma$ on site $i$.
The typical effect of strain on the energy spectrum of electrons is shown in Fig.~\ref{fig1}(c). It is seen that energy bands get flatten. In large period limit, these flat bands near the Dirac point coincide with the Landau levels due to the strain induced pseudo-magnetic fields\cite{pseudoB}. For general periodic perturbation of hopping amplitudes given by Eq.(\ref{tx}), the vector potential associated with
the pseudo-magnetic field is given by $A_x = \sqrt{3} (t_{i_1} -t_{i_2})/2v_F$, $A_y =  (t_{i_1} +t_{i_2}-2t_{i_3})/2v_F$\cite{vectorpotential}, where $t_{i_n}$($n=1,2,3$) are hoppping amplitudes along $\vec{e}_n$ at site $i$ (see Fig.~1a). 

\begin{figure}[t]
\includegraphics[height=1.5in,width=3.4in] {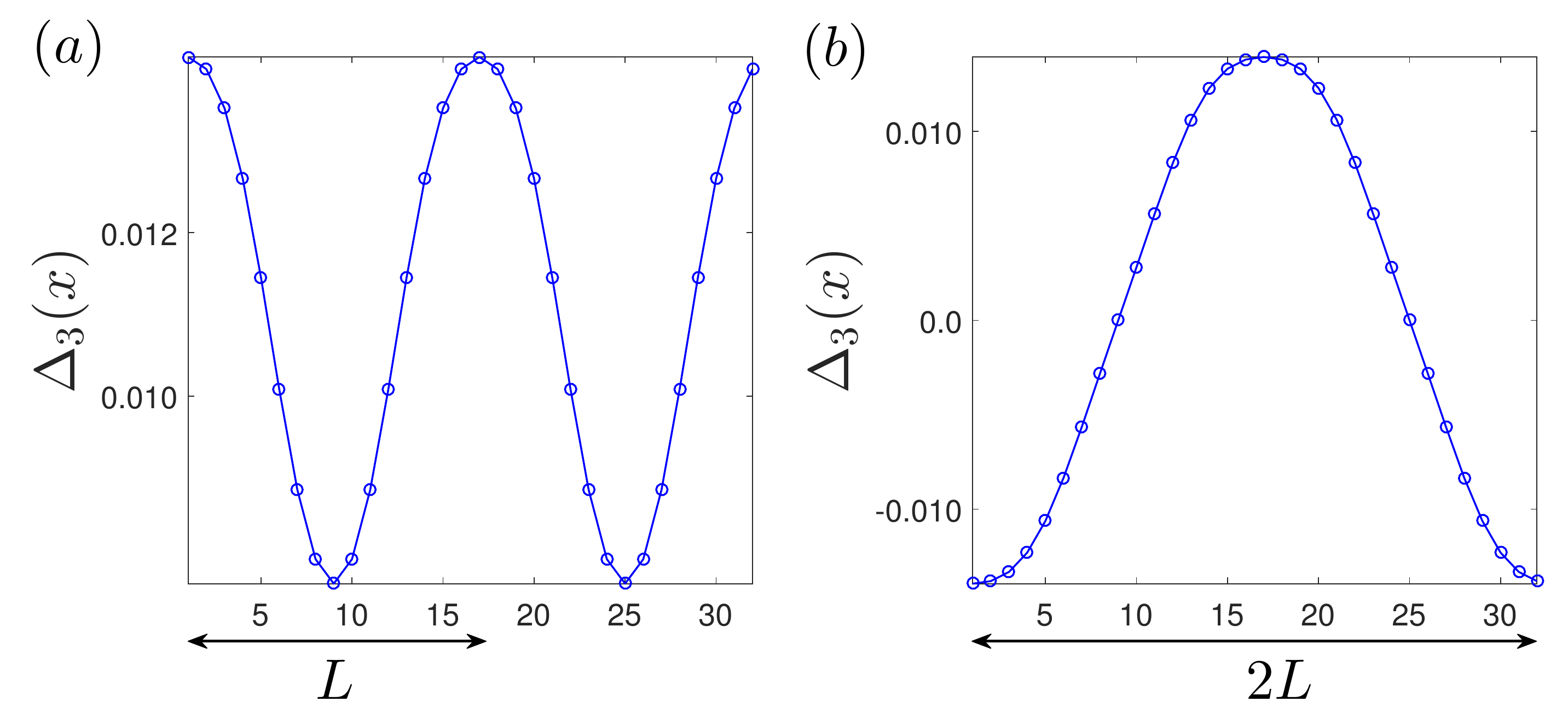}
\caption{Numerical solutions that illustrate the pair density wave with an anomalous period.  Here pairing amplitudes $\Delta_i$ are defined in Fig.~\ref{fig1}, $\alpha =0.025$, and $\delta =0.122$
(a) $\Delta_3$ of the ground state exhibits a uniform order plus a component that oscillates with the same period $L$ of the strain (or period of $L/n$ with $n$ being positive integer). (b) $\Delta_3$ of the metastable state in close to the ground state exhibits anomalous period of $2L$. Here by assuming translational invariance in $y$ direction, mean fields $\chi_{ij}$ and $\Delta_{ij}$ on each bond in real space are solved self-consistently in a $32 \times 32$ lattice with $J/t=1$.}
\label{fig2}
\end{figure}

 Hence for the deformed hopping amplitude of Eq.(\ref{tx}), we have $A_x=0$ and $A_y=-t\alpha \cos (Qx)/v_F$.
The linearize Hamiltonian near K point can be written as 
\begin{equation}
H_q = \hbar v_F
             \left( \begin{array}{cc}
                0  & -i \frac{d}{dx}+i(q_y -A_y)\\
               -i \frac{d}{dx}-i(q_y -A_y) & 0
             \end{array}\right),
\end{equation}
where $q_y$ is the deviation of the wave-vector $\mathbf{k}$ from K. Clearly, for large $L$ (small $Q$),  $H_q$ supports zero energy solutions near $q_y -A_y(x)=0$ with the eigenstate $\psi_0$ being given by
\begin{equation}
\psi _0= N
             \left( \begin{array}{cc}
                \exp^{-\int^x_{x_0} [q_y -A_y(x)]dx} \\ 
               0
             \end{array}\right),   \text{  for $\frac{d }{dx} [q_y -A_y(x)]>0$}
\end{equation}
or 
\begin{equation}
\psi _0= N
             \left( \begin{array}{cc}
                0 \\ 
               \exp^{-\int^x_{x_0} [q_y -A_y(x)]dx} 
             \end{array}\right), \text{ for $\frac{d }{dx} [q_y -A_y(x)]<0$}, 
\end{equation}
where $N$ is a normalization constant and $x_0$ is a root to $q_y -A_y(x)=0$.  It is clear from the above solution that only when $|q_y| \leq t\alpha/v_F$ is satisfied, $x_0$ exists so that zero-energy solutions exist\cite{Jackiw}. This results a flat region along $q_y$ direction ($M$-$K$) as illustrated in Fig~\ref{fig2}(d).

To include correlation effects in flat bands, we consider graphene near half-filling with
the averaged electron density being less than 1. The appropriate model is  
to include the Hubbard interaction between electrons, $
H_U=H_0+U\sum_{i, \sigma}\hat{n}_{i \uparrow}\hat{n}_{i \downarrow}$.
In the strong interacting limit when $\mathbf{U}$ is large, the Hilbert space of the ground state is energetically confined to the singly occupied space described by an effective t-J model given by\cite{tJ}
\begin{eqnarray} 
H= P_G \left[ H_0+ \sum_{\langle ij \rangle} J_{ij} (\vec{S}_{i} \cdot \vec{S}_{j}-\frac{1}{4} n_{i} n_{j}) \right] P_G. \label{tJmodel}
\end{eqnarray}
Here $P_G = \Pi_i (1-n_{i\uparrow} n_{i\downarrow})$ is the Gutzwiller projection operator that projects out states with doubly-occupied sites. $\vec{S}$ and $n$ are spin and number operators for electrons respectively. The antiferromagnetic (AF) coupling, given by  $J_{ij}=4t^2_{ij}/U$, now acquires spatial dependence through
the deformed hopping amplitude $t_{ij}(x_i)$.

To investigate possible phases that arise with the given Hamiltonian $H$, we resort to
the slave-boson method, in which the
no-double-occupancy constraint is implemented  by expressing the electron operator as $c_{i \sigma}=b^{\dagger}_i f_{i \sigma}$ with $b_i$ being the holon carrying the charge and $f_{i\sigma}$ being the spinon carrying the spin\cite{Ubben,sb}. The no-double-occupancy constraint is satisfied by requiring $\sum_{\sigma} f^{\dagger}_{i\sigma} f_{i\sigma} + b^{\dagger}_i b_i=1$. Following Ref.\onlinecite{Ubben}, in the mean-field approximation, $b_i$ is replaced by $\langle b_i \rangle = \sqrt{\delta_i}$ with $\delta_i=1-n_i$ being the hole density at i site. The AF interaction is further decoupled as:  $\vec{S}_{i} \cdot \vec{S}_{j}-\frac{1}{4}\hat{n}_{i}\hat{n}_{j} \rightarrow
-\frac{3}{8} \left(\hat{\chi}^{\dag}_{ij}\hat{\chi}_{ij}+\hat{\Delta}^{\dag}_{ij}\hat{\Delta}_{ij}\right)$, 
where $\hat{\chi}^{\dag}_{ij}=f^{\dag}_{i\uparrow}f_{j\uparrow}+f^{\dag}_{i\downarrow}f_{j\downarrow}$ and $\hat{\Delta}^{\dag}_{ij}=f^{\dag}_{i\uparrow}f^{\dag}_{j\downarrow}-f^{\dag}_{i\downarrow}f^{\dag}_{j\uparrow}$.
Taking the mean-field approximation of the decoupled AF interaction, the mean-field Hamiltonian is given by
\begin{eqnarray} 
&&H_{MF}=\left[\sum_{\langle ij \rangle, \sigma} -\tilde{t}_{ij}f_{i\sigma}^{\dag}f_{j\sigma}
+\sum_{\langle ij \rangle} \tilde{\Delta}^0_{ij}(f^{\dag}_{i\uparrow}f^{\dag}_{j\downarrow}-f^{\dag}_{i\downarrow}f^{\dag}_{j\uparrow})\right]+h.c. \notag \\
&&\qquad \qquad-\sum_{\langle ij \rangle} \tilde{J}_{ij}(|\chi_{ij}|^{2}+|\Delta^0_{ij}|^{2}).
\end{eqnarray}
Here $\chi_{ij}=\langle \hat{\chi}_{ij}\rangle$, $\Delta^0_{ij}=\langle \hat{\Delta}_{ij}\rangle$, 
$\tilde{t}_{ij}=\sqrt{\delta_{i}\delta_{j}}t_{ij}-\tilde{J}_{ij}\chi_{ij}$ is the effective hopping strength,
$\tilde{\Delta}^0_{ij}=\tilde{J}_{ij}\Delta^0_{ij}$, and $\tilde{J}_{ij}=-3J_{ij}/8$. $\chi_{ij}$ and $\Delta_{ij}$ are solved self-consistently through the equations $\chi_{ij}=\langle \hat{\chi}_{ij}\rangle$ and $\Delta^0_{ij}=\langle \hat{\Delta}_{ij}\rangle$
with $\langle \hat{\chi}_{ij}\rangle$ and $\langle \hat{\Delta}^0_{ij}\rangle$ being numerically computed by 
using the mean-field Hamiltonian $H_{MF}$. Note that $\Delta^0_{ij}$ (and thus $\tilde{\Delta}^0_{ij}$ ) is the  average of spinon pairing operator, $\hat{\Delta}_{ij}$, and hence it is not the superconducting amplitude. The superconducting pairing amplitude is the pairing amplitude of of electrons and is given by $\Delta_{ij} = \sqrt{\delta_i \delta_j} \Delta^0_{ij} \approx \delta \Delta^0_{ij}$ with $\tilde{\Delta}_{ij}=\tilde{J}_{ij}\Delta_{ij}$. The superconducting transition temperatures is thus obtained by rescaling the transition temperature for the spinon gap by the average doping $\delta$.
Finally, we note that $H_{MF}$ is essentially the same as the renormalized mean-field Hamiltonian\cite{renormalized} obtained by using the Gutzwiller approximation\cite{Gutzwiller} except that the hopping amplitude $t_{ij}$ and the AF coupling $J_{ij}$ are replaced by $g_t t_{ij}$ and $g_s J_{ij}$ with $g_t = 2 \sqrt{\delta_{i}\delta_{j}}$ and $g_s=4$. Hence both the slave-boson method and the mean-field theory based on the Gutzwiller approximation yields similar results.

To analyze superconducting states in the strain, we define pairing orders on nearest neighboring bonds to any lattice point as shown in Fig.~\ref{fig1}(a) (the same definition applies to $\chi_{ij}$ as well). Note that $\Delta_2 = \Delta_1^*$ is satisfied due to the $C_3$ rotational symmetry.
There are three pairing symmetries in compatible with the symmetry of graphene\cite{symmetry}: extended s-wave, $d_{x^2-y^2}+id_{xy}$, and $d_{x^2-y^2}-id_{xy}$. They can be expressed in terms of pairing amplitudes along three bonds as 
$\Delta_s(x)=\frac{1}{\sqrt{3}}(\Delta_1(x)+\Delta_2(x)+\Delta_3(x))$, $\Delta_{d_{x^2-y^2}}(x)=\frac{1}{\sqrt{6}}(2\Delta_3(x)-\Delta_1(x)-\Delta_2(x))$, and $\Delta_{d_{xy}}(x)=\frac{1}{\sqrt{2}} \text{Im} (\Delta_1(x)-\Delta_2(x))$. In the absence of strain, the
uniform chiral d-wave state,  $d_{x^2-y^2} \pm id_{xy}$, is found to be the superconducting ground state for $J \ge 1$\cite{symmetry}. In the presence of strain, we solve mean-fields $\chi_{ij}$ and $\Delta_{ij}$ on each bond in real space self-consistently.
Fig.~\ref{fig2} (a) and (b) show typical convergent values for $\Delta_3 (x)$.  Due to the imposed periodicity by the strain, one expects that in addition to the uniform $\chi_{ij}=\chi$ and $\Delta_{ij}=\Delta$, $\chi_{ij}$ and $\Delta_{ij}$ of period $L/n$ with $n=1,2,3,...$ (wavevector = $nQ$) are also present and coexist with the uniform orders. This is clearly seen in Fig.~\ref{fig2}(a), in which $\Delta_3$ exhibits period of $L$. However, as indicated in Fig.~\ref{fig2}(b), in addition to period $L$, mean-field orders with anomalous period of $2L$ emerge in certain regime of the strain amplitude $\alpha$.  
\begin{figure}[t]
\includegraphics[height=2.0in,width=2.8in] {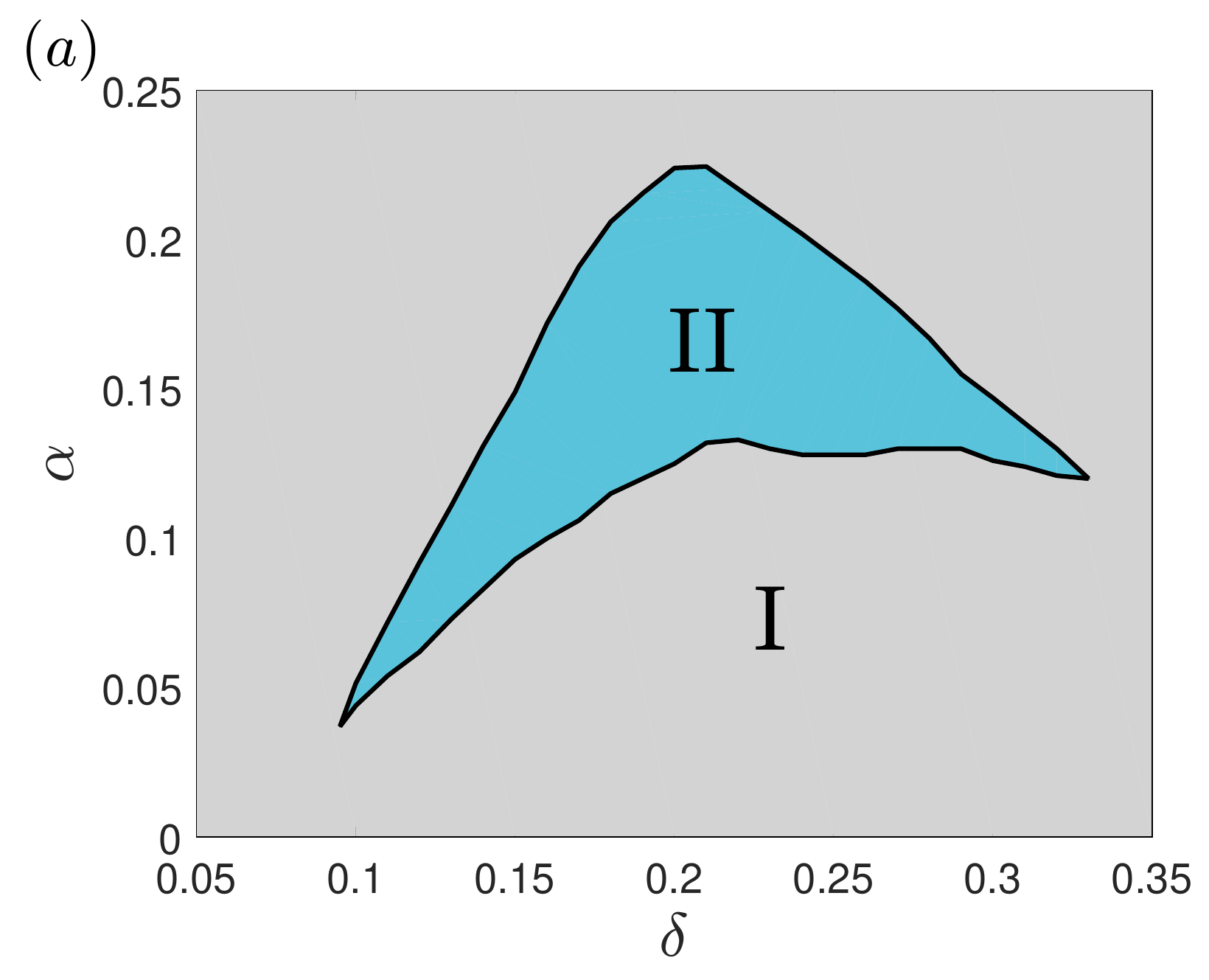}
\includegraphics[height=2.0in,width=2.8in] {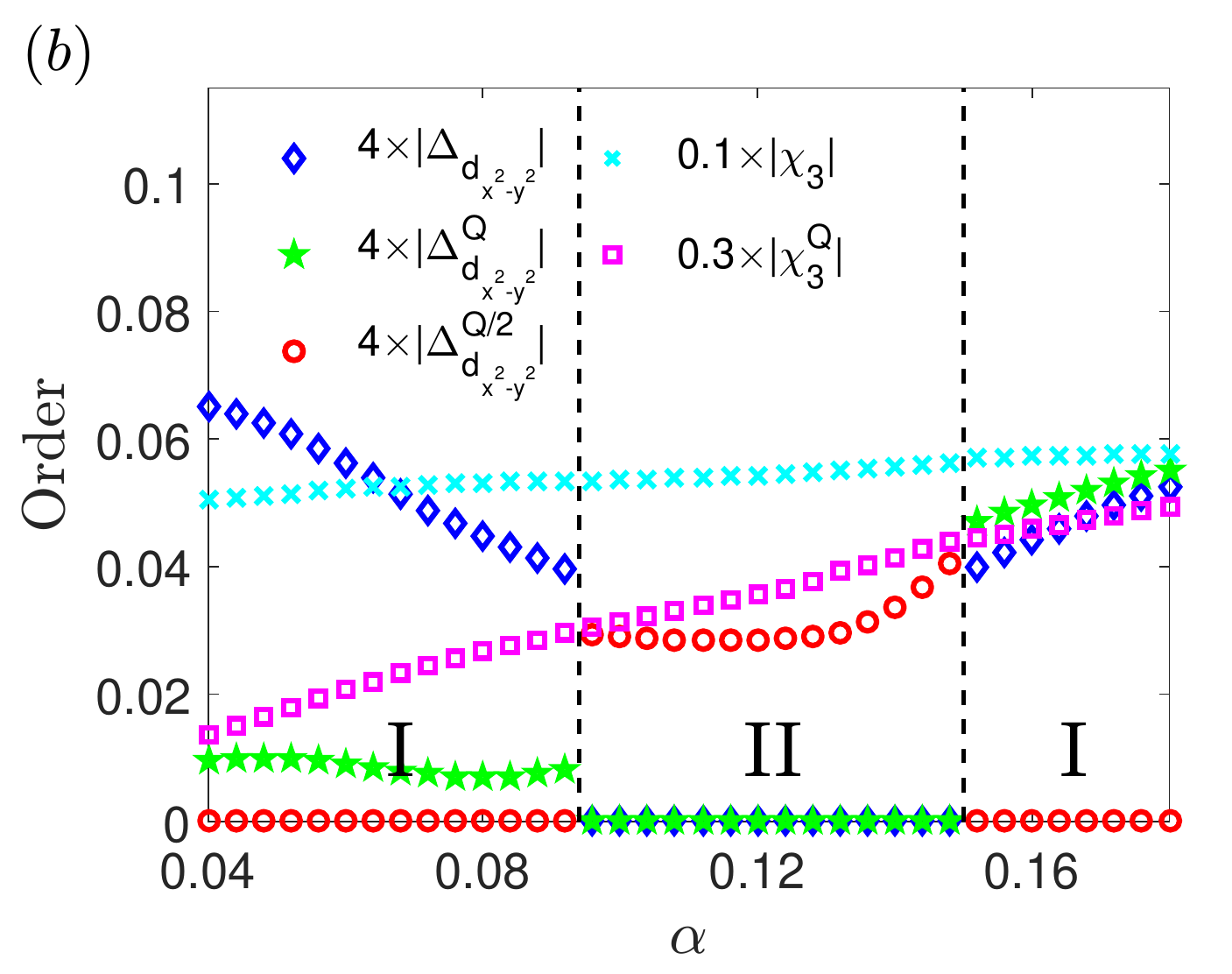}
\caption{ Superconducting phases of strained graphene at zero temperature with $L =16$ and $J/t=1$
(a) Phase diagram in the parameter space of doping $\delta$ and strain $\alpha$. Here in phase I, only orders of integer multiple of wavevector $Q=2\pi/L$, i.e. $nQ$, appear. In phase II, orders with wavevector $Q/2$ coexist with orders with wavevector $nQ$. (b) Quantum phase transitions of mean-field orders for $\delta=0.15$. It is seen that superconducting orders change discontinuously across phase boundaries.}
\label{fig3}
\end{figure}

To further explore the density waves with anomalous period of $2L$, we solve superconducting phases of the graphene
in zero temperature by classifying phases with or without the period of $2L$ (wavevector $Q/2$) as: \\
phase I:  $\Delta_s$, $\Delta_{d_{x^2-y^2} \pm i d_{xy}}$ (uniform orders), $\Delta_s (nQ)$, $\Delta_{d \pm id} (nQ)$, $\chi(nQ)$ and $\rho(nQ)$, \\ 
phase II: $\Delta_{d_{xy}}$, $\Delta_{d_{xy}}(nQ)$, $\chi(nQ)$, $\rho(nQ)$, $\Delta_s(Q/2)$, $\Delta_{d_{x^2-y^2}}(Q/2)$, and $\chi_1(Q/2)$. \\
Here $\rho (nQ)$ represents the on-site charge density wave $\sum_i n_i e^{inQx_i} $ and $\chi(Q/2)$ represents the bond charge density wave $\sum_i \chi_{ij} e^{iQ x_i/2}$. The phase diagram is shown in Fig.~\ref{fig3}(a). It is seen that there is a large region with moderate strain for $\alpha \approx 0.1-0.3$ in which density waves with period of $2L$ (phase II) can be stabilized. For a given doping $\delta$, Fig.~\ref{fig3}(b) shows that as the strain increases, quantum phase transitions occurs with the change of superconducting orders being discontinuously across phase boundaries.

To understand the emergence of orders with wavevector $Q/2$ (period = $2L$), we consider possible
couplings between the charge density wave, the pair density wave and the uniform superconducting order $\Delta$. 
The energy terms in the free energy must conserve the momentum, i.e., the total momentum must vanish. In addition, 
the U(1) symmetry should be respected. As a result, we find that
the lowest order couplings in the free energy are of the form\cite{PDW}: 
$\rho (Q) \Delta^* (Q/2) \Delta (-Q/2)$, $\rho(Q) \Delta^* \Delta (-Q)$, $|\chi (Q/2)|^2 |\Delta(Q)|^2$, and  $|\chi (Q)|^2 |\Delta(Q)|^2$. 
In these lowest coupling terms, the mechanism for the emergency
of finite Cooper pair momentum is due to the momentum conservation. For instance, in the lowest order of the coupling term, $\rho (Q) \Delta^* (Q/2) \Delta (-Q/2)$, the momentum $Q$ carried by CDW is conserved by creating two Cooper pairs with momentum $-Q/2$. The emergent Cooper pair order with momentum $Q/2$ is generally not stable and has to be stabilized as the minimum of the free energy.
For graphene under the strain given by Eq.(\ref{tx}), the induced charge density wave $\rho (Q)$ is proportional to the deformation of hopping amplitude $\delta t \equiv t \alpha$. 
Hence the minimum of the free energy is driven by the couplings
$a(Q) \delta t (Q) \Delta^* (Q/2) \Delta (-Q/2) + b(Q)  |\delta t (Q)|^2 |\Delta(Q)|^2$. Here the coefficients
$a(Q)$ and $b(Q)$ are negative\cite{coefficient} so that both $\Delta(Q/2)$ and $\Delta(Q)$ can be stabilized for sufficiently large $\alpha$.
\begin{figure}[t]
\includegraphics[height=2.4in,width=3.2in] {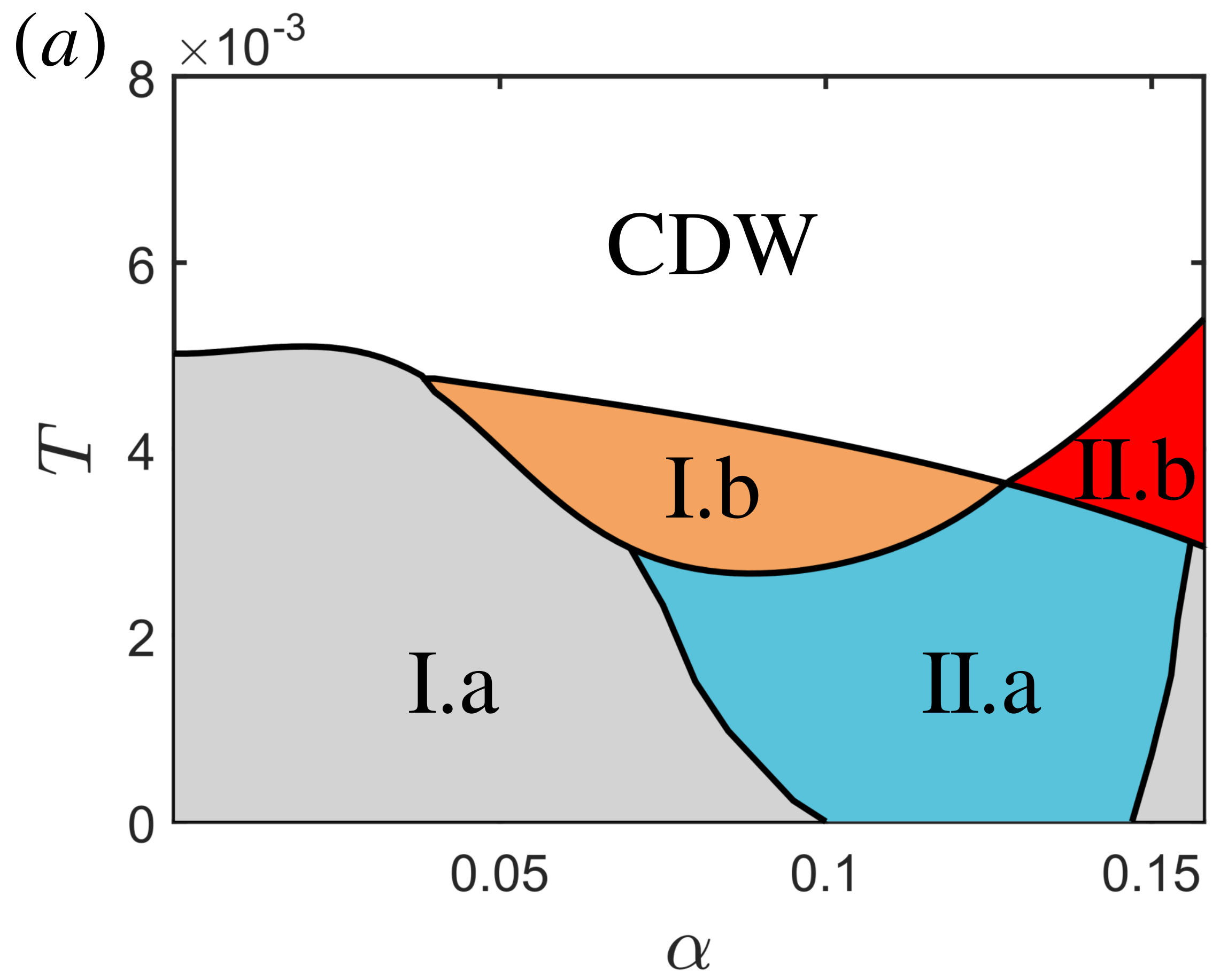}
\includegraphics[height=1.5in,width=1.6in] {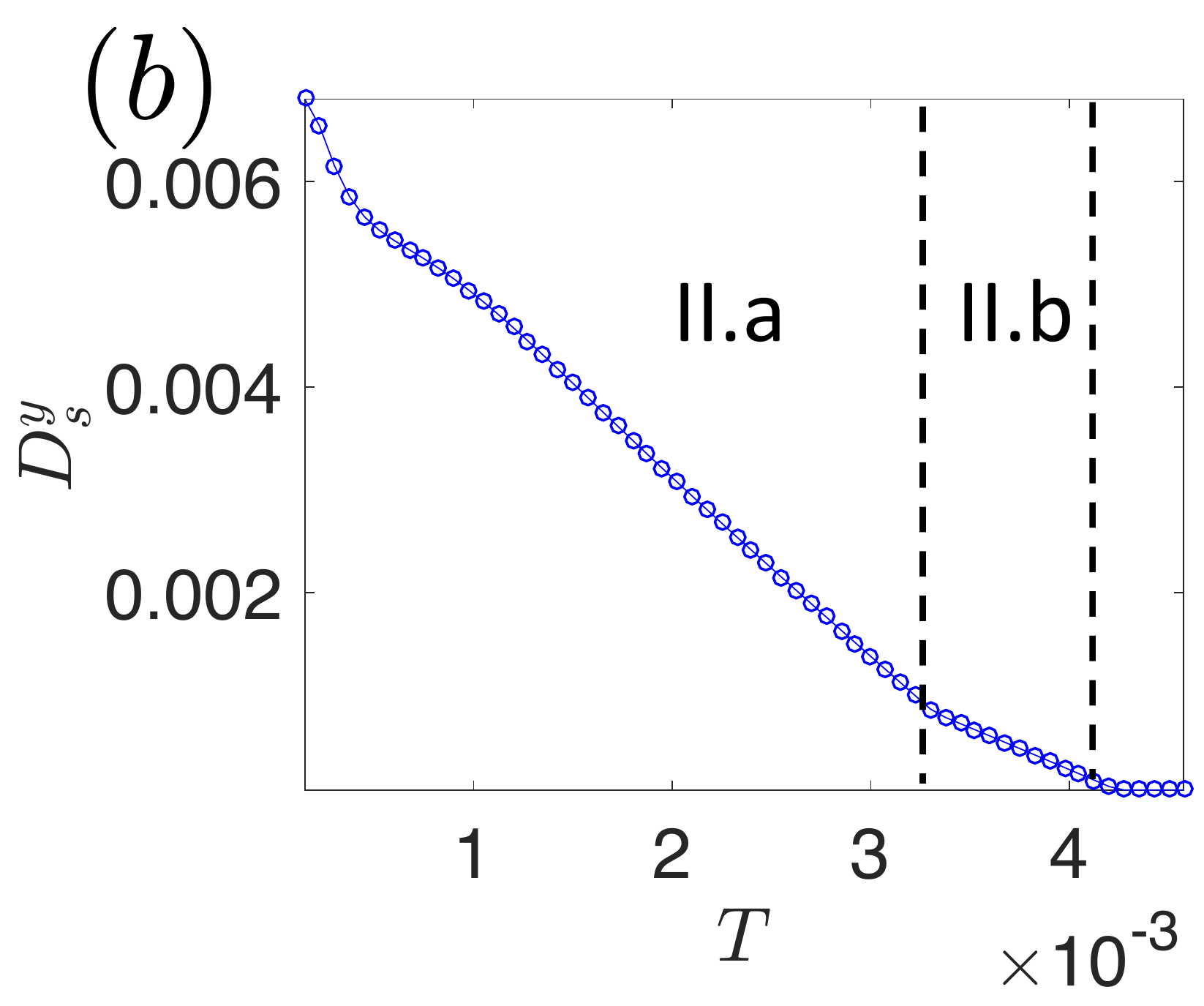}
\includegraphics[height=1.6in,width=1.6in] {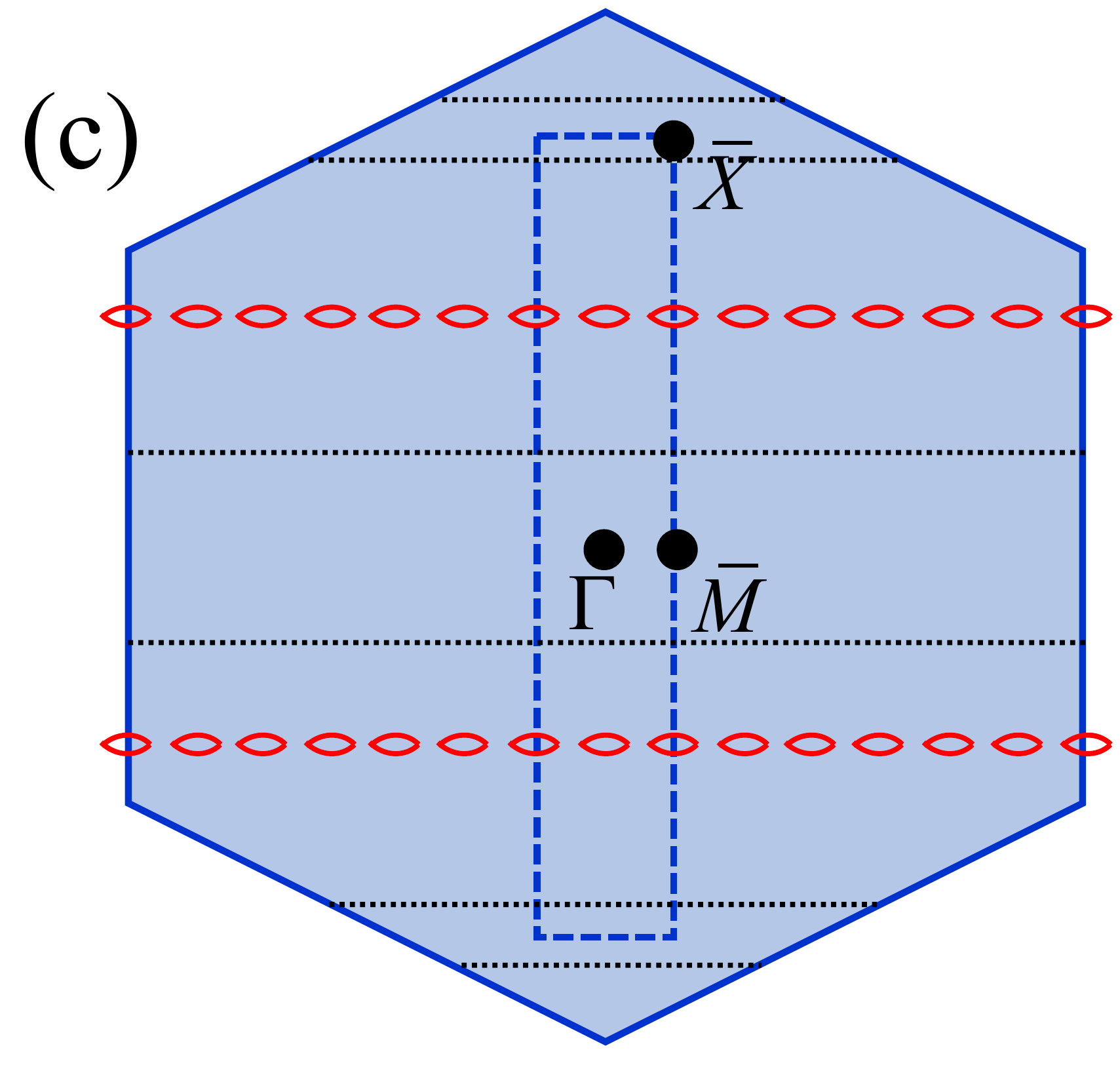}
\caption{Superconducting phases of strained graphene at finite temperatures with $L =16$, $J/t=1$ and $\delta=0.15$
(a) Distinct phases with different superconducting orders (see text for more details) at finite temperatures. Here phase II.b is a pure superconducting PDW state with  non-zero center-of-mass momentum for Cooper pairs and coexists with the CDW order. (b) Superfluidity weight  along a cut from phase II.a to the CDW phase with $\alpha=0.14$. The non-vanishing $D_y$ implies that the pure PDW state in phase II.b is superconducting.(c) Nodal rings (indicated by red color) of quasi-particles in phase II.b. In addition to nodal rings, flat-bands marked by black solid lines are also on the Fermi surface  in the normal states without pairing density waves.}
\label{fig4}
\end{figure}
However, due to different dependence on $\alpha$, $\Delta(Q/2)$ and $\Delta(Q)$ compete with each other and eventually $\Delta(Q/2)$ wins, resulting in the emergence of phase II as an intermediate phase.

\begin{figure}[t]
\includegraphics[height=2.2in,width=3.0in] {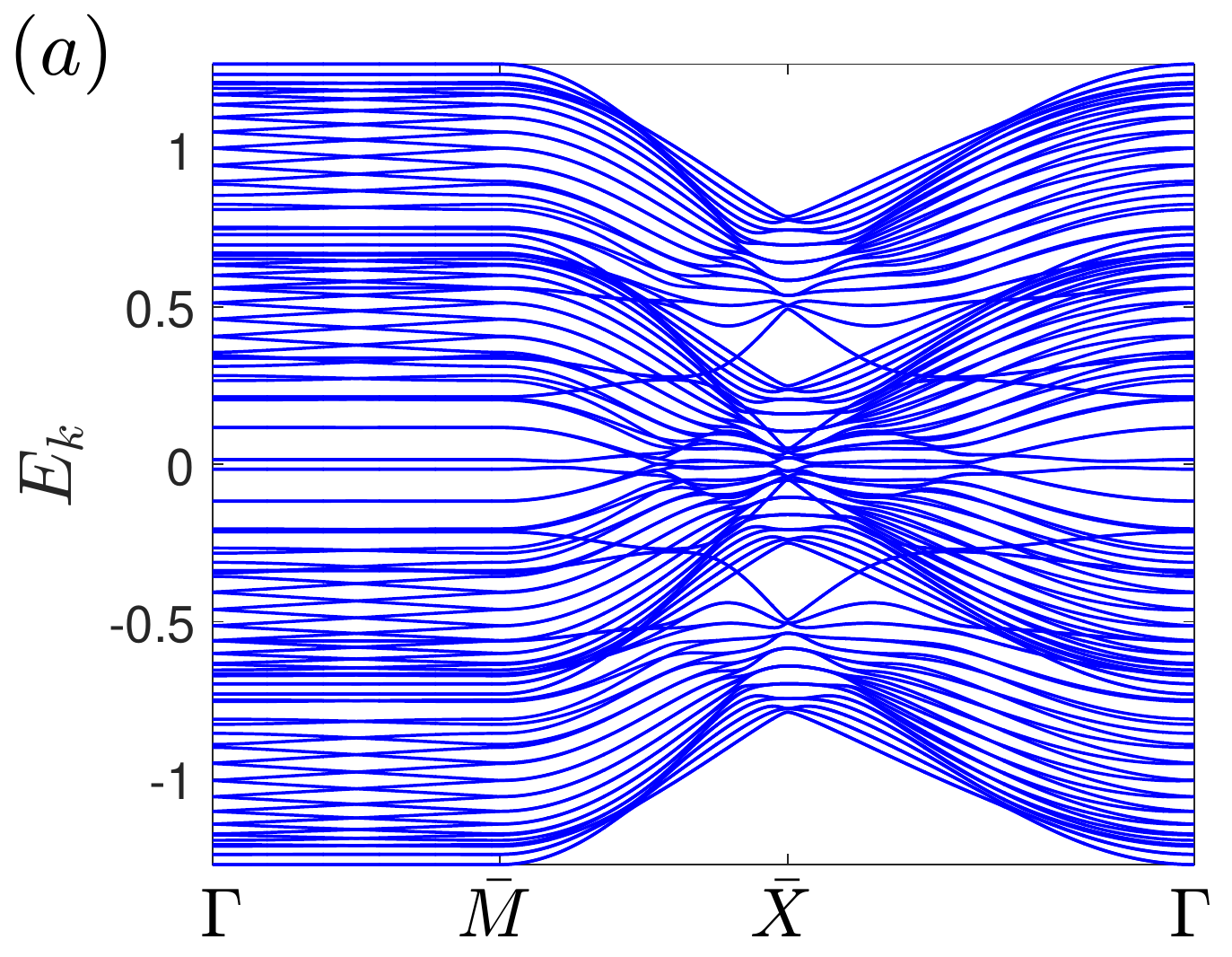}
\includegraphics[height=2.2in,width=3.0in] {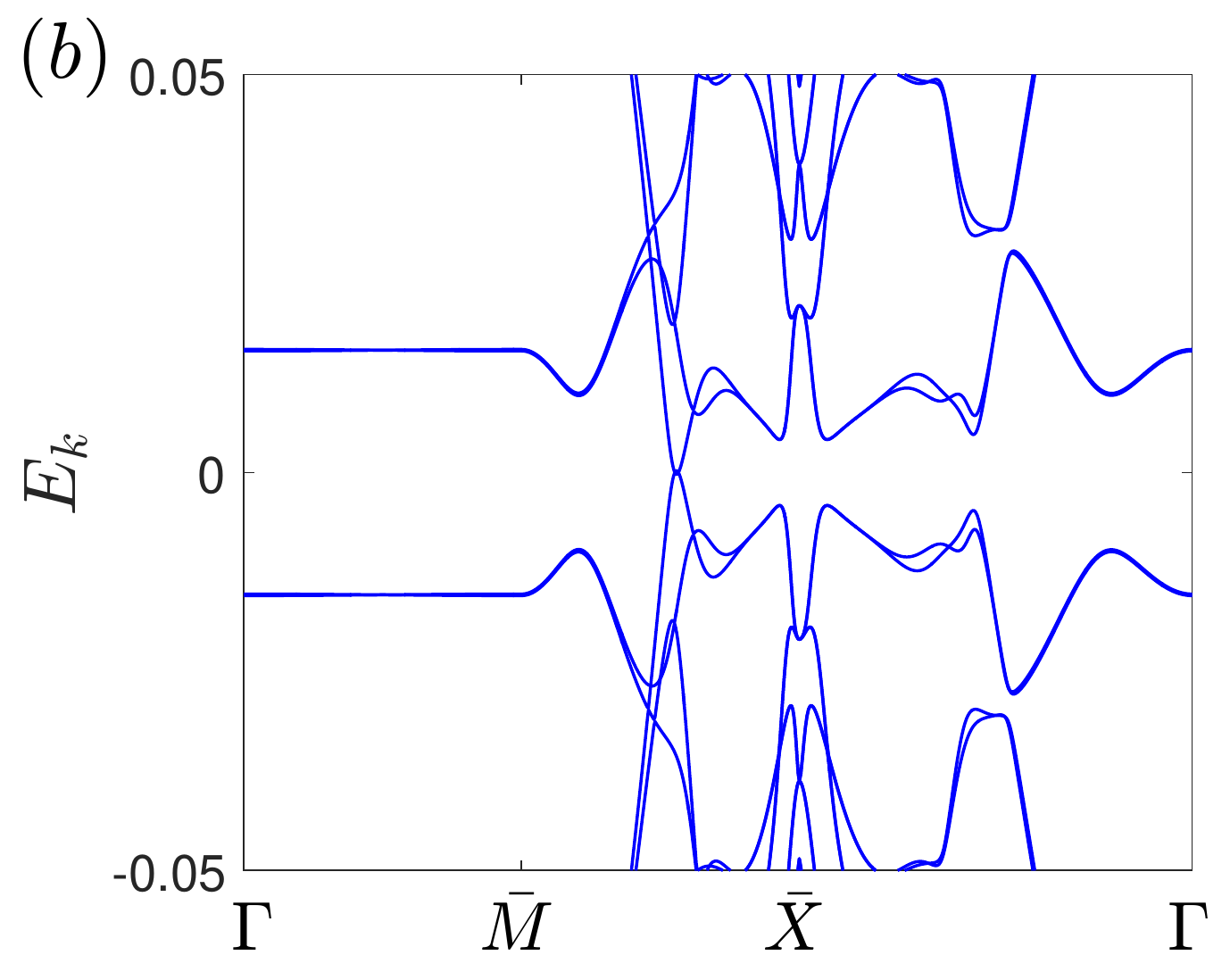}
\caption{(a) Quasi-particle excitations in phase II.b. (b) Blown-up of the energy spectrum near $E_k \sim 0$ shown in (a). Here $\bar{M}$ and $\bar{X}$ are points at the boundary of reduced Brillouin zone shown in Fig.4(c) in the text. It is seen that $E_k$ only vanishes at nodal points (from the nodal rings) in going from  $\bar{M}$ to $\bar{X}$.} \label{fig5}
\end{figure}
At finite temperatures, the competition of different superconducting orders lead to more complicated phase diagram as shown in Fig.~\ref{fig4}(a).  Here orders emerging in different phases are: \\
phase I.a: $\Delta_s$, $\Delta_{d \pm i d }$, $\Delta_s (nQ)$, $\Delta_{d \pm id} (nQ)$, $\chi(nQ)$, and $\rho(nQ)$, \\ 
phase I.b: $\Delta_{d_{xy}}$,  $\Delta_{d_{xy}} (nQ)$, $\chi(nQ)$, and $\rho (nQ)$, \\ 
phase II.a: $\Delta_{d_{xy}}$, $\Delta_{d_{xy}}(nQ)$, $\chi(nQ)$, $\rho(nQ)$, $\Delta_s (Q/2)$, $\Delta_{d_{x^2-y^2}} (Q/2)$, and $\chi_1 (Q/2)$, \\ 
phase II.b: $\rho(nQ)$, $\chi(nQ)$, $\Delta_s (Q/2)$, and $\Delta_{d_{x^2-y^2}} (Q/2)$. \\ 
Here for small $\alpha$, when going from phase I.a to phase I.b, superconducting order $\Delta_i$ become pure imaginary and only $\Delta_1$ and $\Delta_2$ survive, while for large $\alpha$, going from phase II.a to phase II.b, $\Delta_{d_{xy}}(Q)$ and $\chi_1(Q/2)$ disappear. The driving coupling for disappearance of $\Delta_{d_{xy}}(Q)$ and $\chi_1(Q/2)$ is the coupling $|\chi (Q/2)|^2 |\Delta(Q)|^2$. Remarkably, due to this coupling, we see that a pure PDW state that coexists with the CDW order emerges at some finite temperature with moderate strain $\alpha \gtrsim 0.125$ (phase II.b). Furthermore, as shown in Fig.~\ref{fig4}(b), by computing the superfluid weight \cite{superfluid_density}, we find that phase II.b is superconducting, in contrast to the CDW state with vanishing superfluid weight. 
Similar to the PDW state observed in high Tc cuprates in which the quasi-particle excitations are gapless with Fermi arcs displayed at finite temperatures\cite{PALee}, here the phase II.b is also gapless with nodal rings (red curves) as shown in Fig.~\ref{fig4}(c). The exact location of the nodal ring can be exhibited in the corresponding energy spectrum, which is plotted in Figs.~\ref{fig5}(a) and (b), showing the energy spectrum of the quasi-particle excitation along the path $\Gamma$-$\bar{M}$-$\bar{X}$-$\Gamma$. Here the blown-up of Fig.~\ref{fig5}(a) for $E_k \sim 0$  is shown in Fig.~\ref{fig5}(b), indicating the location of  the nodal ring in going from $\bar{M}$ to $\bar{X}$.

Note that without flat-bands, it is generally more difficult to have pair of states near the Fermi surface to satisfy the condition: total momentum is $Q/2$. Hence density for pair of states near the Fermi surface with total momentum $Q/2$ is low. In the presence of flat bands, it is much easier to satisfy the condition with the total momentum being $Q/2$ as the energy does not depend on the momentum. Therefore, flat-bands help in stabilizing the Cooper pair with momentum $Q/2$. This is illustrated in  Fig.~\ref{fig4}(c), which shows flat-bands (black solid lines) on the Fermi surface in the normal state are gapped out due to pairing of electrons with center of mass momentum being $Q/2$, while the same pairing is not possible for ring-shape Fermi surfaces, leaving nodal rings as gapless excitations in phase II.b.
Phase II.b is thus a unique realization of the long searched superconducting state with non-vanishing center of mass momentum for Cooper pairs.

\section{Discussion and Summary}
In summary, while superconductivity is discovered to be realized in a flat-band in a twisted bi-layer graphene, we find that the same chiral d-wave superconductivity can be also realized in topological flat-bands induced by strain in graphene through periodic ripples. The stabilization of chiral d-wave superconductivity is through the enhancement of the correlation effect in flat bands. As a result, even for slightly doped graphene, the graphene can be turned into a chiral d-wave superconductor by applying strain. The uniform chiral d-wave superconductivity generally coexists with the CDW order and chiral PDW order. At finite temperatures, it is further found that a pure superconducting PDW state with coexisting CDW emerges in graphene under moderate strain strength. The emergent pure superconducting PDW state is the realization of  the long searched superconducting state with non-vanishing center of mass momentum for Cooper pairs.  

Finally, we discuss feasibility of realizing the superconducting PDW state 
and the experimental features that can be observed. First, distinguishing
the superconducting PDW state from other superconducting state can be generally
detected by using the scanning tunneling microscope. One expects that the energy gap observed in the 
differential conductance measurement depends on the position and exhibits oscillatory behavior.
For the feasibility of realizing the superconducting PDW state, so far our analysis has focused on 
nanoscale ripples (wavelength from 0.1 nm to 10nm), which have
been observed experimentally\cite{strain,strain2}. It is known that 
the generation of flat-bands by ripple depends on the ratio of the height $h$ to the period $L$.
When the condition $h^2/La \ge 1$  is met, flat-bands arise\cite{strain2}. Since $\alpha$ that characterizes
the deformation of hopping amplitude depends only on $h/L$, for a given $\alpha$, increasing height of the ripple would generate
flat-bands for micron-size ripples. Hence our results are also applicable to micron-size ripples.
For ripples of micron-size, the Cooper pair momentum $Q/2$ is smaller. Furthermore, since the energy barrier for realizing
the superconducting PDW state is essentially the kinetic energy of the Cooper pair  with momentum being
$Q/2$, we expect that the energy barrier for realizing the PDW state is lower for ripples of micron-size.
It is therefore easier to realize the superconducting PDW state in micron-size ripples.

The optimal strain needed to realize the PDW state can be read off from Fig.~\ref{fig4}(a) with
$\alpha \sim 0.14$, with the corresponding aspect ratio of the ripple being $L/h \approx 20$. 
The minimum height $h$ thus needs to satisfy $h/L \ge 0.05$. Together with the requirement $h^2/La \ge 1$,
the height requires to realize the PDW state is $h \ge 4 a$ for $L=16a$, which can be engineered by appropriate 
choosing misfit of the thermal expansion between graphene and the substrate\cite{strain}.  
On the other hand, the critical temperature for accessing the PDW state is around $1$ meV (a few K) for $J/t=1$ and is expected
to be further reduced for micron-size ripples. Our analyses thus indicate that it is feasible experimentally to realize
the long searched superconducting state with non-vanishing center of mass momentum for Cooper pairs. Therefore,  
results of this work illustrate the feasibility for graphene under strain to be a tunable platform for realizing both novel superconducting 
orders and charge density wave orders.

\section*{Acknowledgement}
We acknowledge support from the Ministry of Science and
Technology (MoST), Taiwan. In addition, we also acknowledge support from Center for Quantum Technology,
TCECM, and Academia Sinica Research Program on Nanoscience and Nanotechnology, Taiwan.

\end{document}